\documentclass[letterpaper]{article} 
\usepackage{aaai2027}  
\usepackage[hyphens]{url}  
\usepackage{graphicx} 
\usepackage{natbib}  
\usepackage{caption} 
\usepackage{algorithm}
\usepackage{algorithmic}
\usepackage{amsmath}
\usepackage{amssymb}
\usepackage{array}
\usepackage{tabularx}
\usepackage{multirow}

\usepackage{newfloat}
\usepackage{listings}
\DeclareCaptionStyle{ruled}{labelfont=normalfont,labelsep=colon,strut=off} 
\floatstyle{ruled}
\newfloat{listing}{tb}{lst}{}
\floatname{listing}{Listing}

\usepackage{booktabs}

\title{From Script to Drama: An Agentic Framework for Controllable Multi-Speaker Dialogue TTS}
\author{
    Kangxiang Xia\textsuperscript{\rm 1},
    Xinfa Zhu\textsuperscript{\rm 1},
    HangRui Hu\textsuperscript{\rm 2},
    Kexin Huang\textsuperscript{\rm 2},\\
    Wenjie Tian\textsuperscript{\rm 1},
    Ziyue Jiang\textsuperscript{\rm 2},
    Bingshen Mu\textsuperscript{\rm 1},
    Jingbin Hu\textsuperscript{\rm 1},\\
    Ting He\textsuperscript{\rm 2},
    Lei Xie\textsuperscript{\rm 1}\corresponding,
    Jin Xu\textsuperscript{\rm 2}
}
\affiliations{
    \textsuperscript{\rm 1}Audio, Speech and Language Processing Group (ASLP@NPU), Northwestern Polytechnical University, Xi'an\\
    \textsuperscript{\rm 2}Independent Researcher\\
    xkx@mail.nwpu.edu.cn, lxie@nwpu.edu.cn\\
    Demo page: \url{https://xkx-hub.github.io/FStD-demo/}
}

\nocopyright

\begin{document}

\maketitle

\begin{abstract}

Multi-speaker dialogue TTS requires natural speech generation, consistent speaker identity, coherent cross-turn transitions, and fine-grained control of expressive attributes such as emotion, speaking rate, and loudness. These requirements are difficult to satisfy reliably with one-shot generation, especially in long-form dialogue.
We propose a controllable multi-speaker dialogue TTS framework that formulates synthesis as critique-driven iterative refinement. Its speech backbone, ControlEdit-TTS, unifies instruction-following synthesis and natural-language-guided attribute editing, enabling correction of expressive errors without full regeneration. The framework further performs hierarchical utterance-level and scene-level critique, routing detected issues to editing, resynthesis, or timing adjustment.
Experiments on a bilingual Chinese--English dialogue benchmark show improved utterance-level instruction following, better dialogue-level preference than direct dialogue models and agentic baselines, and more effective refinement than regeneration-only alternatives while preserving speaker identity. Ablations further confirm the benefits of scene-level critique and edit-based correction.
\end{abstract}

\section{Introduction}

Recent advances in neural text-to-speech (TTS) have greatly improved the naturalness and expressiveness of synthetic speech~\cite{f5tts,ns2,valle,kalle,Qwen3-TTS,Cosyvoice2}. Beyond single-speaker reading-style synthesis, there is growing interest in multi-speaker dialogue generation for applications such as dubbing, audiobooks, podcasts, and audio drama~\cite{soulxpodcast,vibevoice,AuDirector,dopamine}. In these settings, a system must do more than generate intelligible speech for multiple characters. It must preserve speaker identity across turns, produce coherent transitions between speakers, and realize utterance-level expressive intent through attributes such as emotion, speaking rate, and loudness. These requirements are especially demanding in long-form dialogue, where both utterance quality and scene-level coherence must be maintained.

\begin{figure}[tbp]
    \centering
    \includegraphics[width=\linewidth]{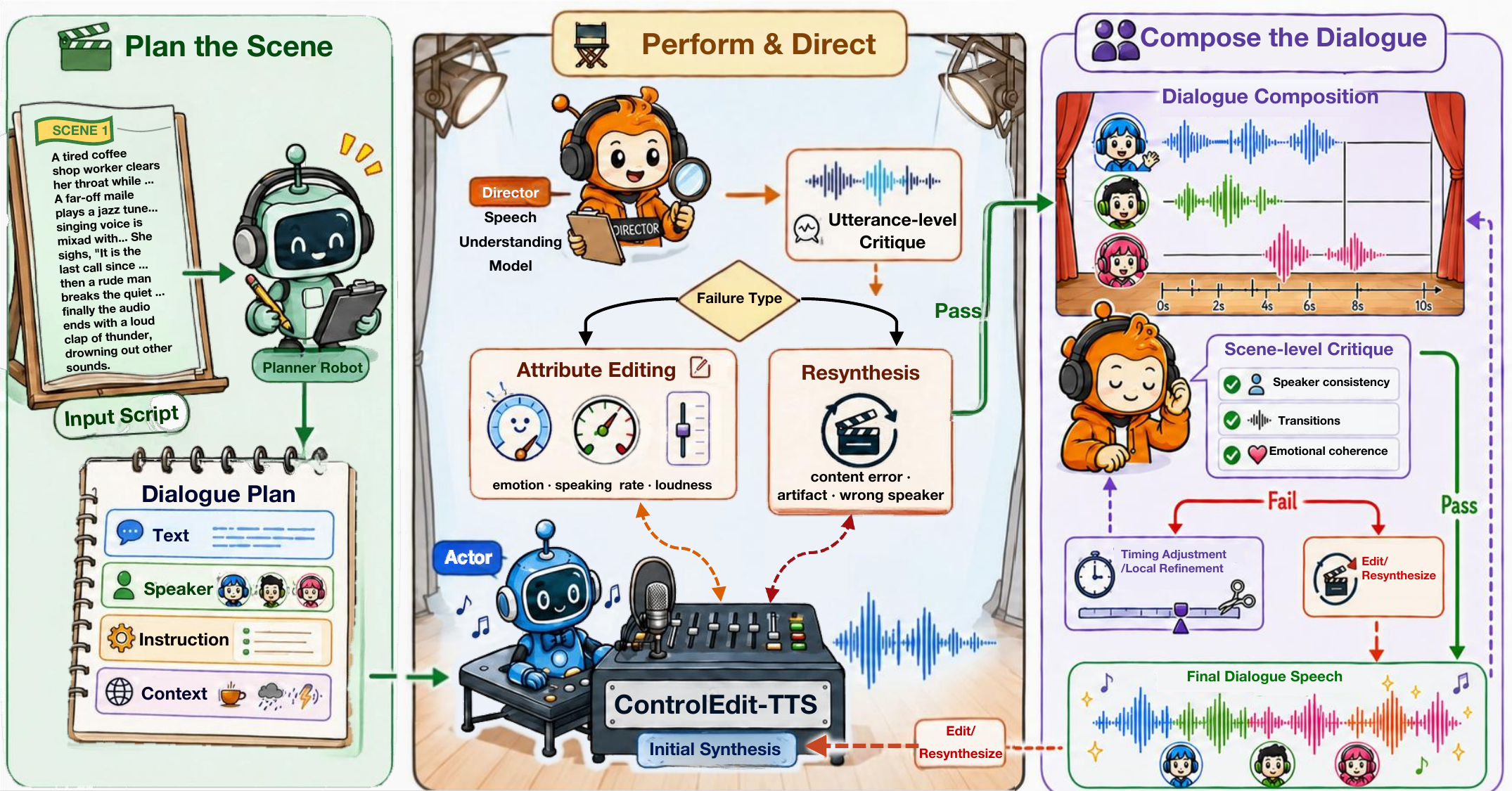}
    \caption{Overview of our method.}
    \label{fig:over}
    \vspace{-20pt}
\end{figure}

Existing work addresses these requirements only partially. Direct multi-speaker dialogue systems can model long-form conversations and multiple roles~\cite{CoVoMix,CoVoMix2,vibevoice,JoyTTS,AuDirector}, often producing globally coherent multi-turn outputs, but they typically provide limited or inflexible control over how each utterance should be performed. In parallel, instruction-following TTS models~\cite{voxcpm2,Qwen3-TTS,Voicesculptor,Cosyvoice3,OV-InstructTTS} can follow natural language descriptions of speaking style, such as emotion, pacing, and vocal energy, for arbitrary prompt voices, enabling much finer utterance-level expressiveness. However, they are primarily designed for isolated utterances and do not explicitly address dialogue-level consistency, cross-turn coherence, or context-aware correction in multi-character settings. As a result, current systems still struggle to provide both fine-grained utterance-level controllability and robust dialogue-level quality within a unified generation framework.

We argue that this limitation stems in part from how the problem is formulated. Controllable dialogue TTS is still treated largely as a one-shot generation task: given a script and style requirement, the model produces each utterance once and hopes the result is correct. In practice, however, many failures are local rather than global. An utterance may preserve the correct text and speaker identity yet under-express the requested emotion, sound slightly too fast, or exhibit loudness that is inconsistent with the dramatic context. Full resynthesis is a coarse remedy for such errors, and repeated resynthesis becomes increasingly inefficient and unstable in long-form production workflows~\cite{Audio-Oscar}. Moreover, utterances that are individually acceptable may still produce abrupt transitions or inconsistent expressive progression once assembled into a dialogue. These observations suggest two requirements for controllable dialogue synthesis: the system should be able to revise local expressive errors without unnecessarily discarding acceptable speech, and it should evaluate not only individual utterances but also the composed scene.

Based on this view, we formulate controllable multi-speaker dialogue TTS as a critique-driven iterative refinement problem rather than one-shot generation. We propose an agentic framework that decomposes dialogue synthesis into four stages: planning, synthesis, critique, and targeted revision. Given a script, character information, and utterance-level control instructions, the framework first constructs a structured dialogue plan and synthesizes the dialogue turn by turn. 
It then performs hierarchical critique at both the utterance and scene levels, checking local instruction adherence and synthesis quality as well as cross-turn consistency and contextual appropriateness over the composed dialogue. 
Crucially, the resulting feedback is converted into executable actions: editable expressive mismatches are corrected through speech editing, severe failures trigger selective resynthesis, and temporal discontinuities are handled through timing adjustment. This action-oriented design allows the system to refine only problematic regions while preserving acceptable speech elsewhere.

At the core of the framework is ControlEdit-TTS, a unified speech model that supports both instruction-following synthesis and natural-language-guided expressive editing within the same autoregressive architecture. Given an existing utterance and corrective feedback, ControlEdit-TTS can modify three practically important attributes---emotion, speaking rate, and loudness---while preserving linguistic content and speaker identity as much as possible. This unified design is important for iterative dialogue refinement: many failures in controllable dialogue synthesis are not complete synthesis breakdowns, but localized expressive mismatches that are better handled through editing than through full resynthesis. By supporting both initial generation and targeted correction within the same model, ControlEdit-TTS provides a natural backbone for critique-guided dialogue refinement.

To support systematic evaluation, we further construct a bilingual Chinese--English benchmark for controllable multi-speaker dialogue TTS and establish a multi-level evaluation protocol covering utterance-level instruction adherence and speaker preservation, dialogue-level preference, and refinement efficiency. Experiments show that the proposed approach improves utterance-level controllability while preserving speaker identity, achieves stronger dialogue-level preference than both direct dialogue models and agentic baselines, and reduces residual errors more effectively than resynthesis-only refinement. These results highlight the value of combining hierarchical dialogue-aware critique with a unified speech model for synthesis and editing.

In summary, our contributions are threefold. First, we formulate controllable multi-speaker dialogue TTS as an iterative refinement problem and propose a hierarchical plan--synthesize--critique--refine framework with both utterance-level and scene-level quality control. Second, we develop ControlEdit-TTS, a unified model for instruction-following synthesis and attribute-level editing of emotion, speaking rate, and loudness. Third, we construct a bilingual benchmark and a multi-level evaluation protocol for controllable dialogue speech generation. Experiments show improved controllability, stronger dialogue-level preference, and more effective refinement than resynthesis-only alternatives.

\section{Related Work}

\subsection{Instruction-following and Editable TTS}

Controllable TTS has evolved from predefined style labels and reference speech toward natural-language descriptions. PromptTTS~\cite{Prompttts}, InstructTTS~\cite{InstructTTS}, VoxInstruct~\cite{VoxInstruct}, ControlSpeech~\cite{Controlspeech}, Qwen3-TTS~\cite{Qwen3-TTS}, and VoiceSculptor~\cite{Voicesculptor} substantially improve the flexibility of language-based style control. However, most of these systems still formulate instruction-following synthesis as one-shot generation, so local expressive failures are typically handled by full resynthesis.

A related line of work studies unified speech generation and editing. Voicebox~\cite{Voicebox}, SpeechX~\cite{SpeechX}, and VoiceCraft~\cite{VoiceCraft} show that synthesis and editing can be supported within general speech generation models. However, existing editing methods mainly focus on content replacement, restoration, or independently specified transformations, rather than natural-language-guided expressive correction within an automatic critique-and-refine loop. In contrast, our ControlEdit-TTS unifies instruction-following synthesis and expressive attribute editing in the same model for iterative dialogue refinement.

\subsection{Agentic Systems for Audio Generation}

Agentic systems improve complex tasks through planning, feedback, and revision. ReAct~\cite{React}, Self-Refine~\cite{Self-refine}, and Reflexion~\cite{Reflexion} establish these mechanisms in language tasks, but applying them to speech generation additionally requires interpreting acoustic failures and mapping them to executable synthesis operations.

Several studies extend LLM orchestration to speech and audio generation. AudioGPT~\cite{Audiogpt}, WavJourney~\cite{Wavjourney}, PodAgent~\cite{Podagent}, Dopamine Audiobook~\cite{dopamine}, AuDirector~\cite{AuDirector}, and Audio-Oscar~\cite{Audio-Oscar} demonstrate the value of planning and feedback in long-form audio production. However, their refinement typically operates at the workflow, prompt, event, or mix level, rather than diagnosing utterance-level expressive errors and routing them to editing or resynthesis within a unified TTS model. Our framework instead performs hierarchical critique over utterances and composed dialogues, and converts the diagnosed failure type into executable editing, resynthesis, or timing-adjustment actions.

\section{Method}

\begin{figure*}[htbp]   
    \centering
    \includegraphics[width=\textwidth]{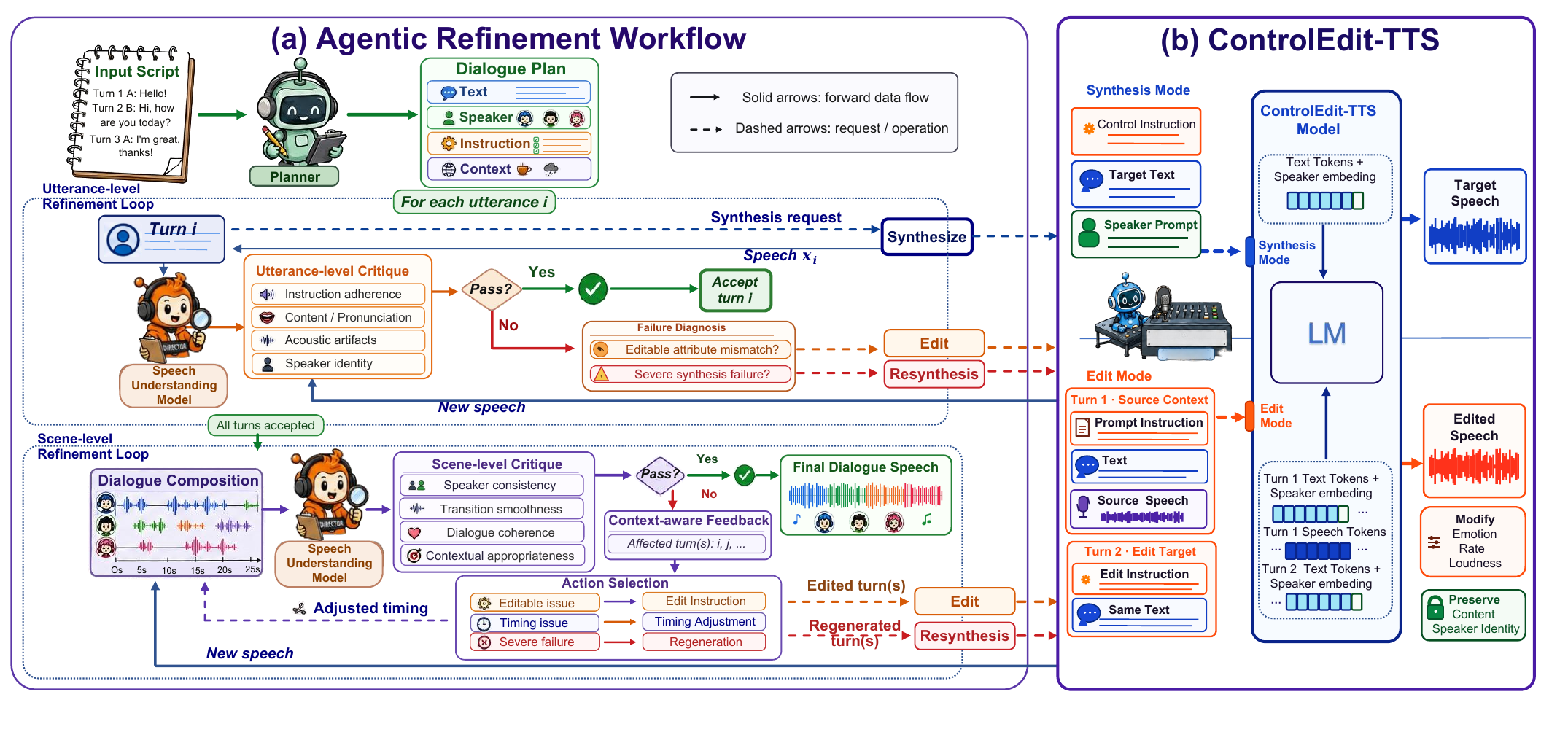}
    \caption{Overview of the proposed method. 
    (a) The agentic refinement framework converts an input script into a structured dialogue plan and iteratively applies utterance- and scene-level critique to trigger attribute editing, selective resynthesis, or timing adjustment. 
    (b) ControlEdit-TTS unifies instruction-following synthesis and attribute-level editing within a shared autoregressive model, enabling targeted correction of emotion, speaking rate, and energy/loudness while preserving linguistic content and speaker identity.}
    \label{fig:method_overview}
\end{figure*}

Our method formulates controllable multi-speaker dialogue TTS as a closed-loop refinement process. Given a script, the framework constructs a structured dialogue plan, synthesizes each turn, critiques both utterance-level instruction adherence and scene-level consistency, and maps detected issues to editing, resynthesis, or timing adjustment. At the core of the framework is ControlEdit-TTS, a unified model built on Qwen3-TTS-12Hz-1.7B-CustomVoice~\cite{Qwen3-TTS} that supports both instruction-following synthesis and attribute-level editing of emotion, speaking rate, and loudness. The following sections describe the unified synthesis-edit model and the hierarchical refinement framework.

\subsection{ControlEdit-TTS: Unified Synthesis and Editing}

ControlEdit-TTS is a unified autoregressive speech model for both instruction-following synthesis and attribute-level editing. As illustrated in Figure~\ref{fig:method_overview}(b), the two modes share the same backbone and speaker-conditioning mechanism and differ only in their input organization. We focus on three expressive attributes---emotion, speaking rate, and loudness---and perform targeted modification of these attributes while preserving linguistic content and speaker identity.

\subsubsection{Task Formulation}

Let $G_{\theta}$ denote ControlEdit-TTS, $t$ the target text, $c$ a natural-language control instruction, and $s$ the speaker condition derived from a reference voice. In synthesis mode, the model directly generates speech according to the text, speaker identity, and expressive instruction:
\begin{equation}
x = G_{\theta}(t,c,s;\mathrm{syn}),
\end{equation}
where $x$ denotes the synthesized speech and $c$ specifies the target expressive condition, e.g., angry, slower, or quieter.

In edit mode, the model additionally receives a source speech segment $x$, its original control condition $c_{\mathrm{src}}$, and a target edit instruction $c_{\mathrm{edit}}$. The operation is formulated as
\begin{equation}
x' =
G_{\theta}(x,t,c_{\mathrm{src}},c_{\mathrm{edit}},s;\mathrm{edit}),
\end{equation}
where $x'$ is the edited speech. Here $t$ and $s$ remain unchanged, while $c_{\mathrm{edit}}$ specifies the desired expressive modification, such as making the utterance more angry, slightly faster, or quieter. The current formulation is restricted to expressive-attribute editing and does not perform content modification or speaker-identity conversion.

\subsubsection{Unified Autoregressive Modeling}

ControlEdit-TTS is initialized from Qwen3-TTS-12Hz-1.7B-CustomVoice~\cite{Qwen3-TTS}. We retain its architecture, speech tokenizer, and autoregressive audio-token modeling formulation, and extend the training task to support both synthesis and editing. In synthesis mode, the model predicts target speech tokens from the control instruction, text, and speaker condition. In edit mode, it receives a two-turn sequence consisting of the source instruction, text, and speech tokens, followed by the target edit instruction and repeated text under the same speaker condition. The edit context is
\begin{equation}
\mathcal{C}_{\mathrm{edit}} =
\bigl((c_{\mathrm{src}}, t, x), (c_{\mathrm{edit}}, t), s\bigr),
\end{equation}
from which the model predicts the edited speech $x' = G_{\theta}(\mathcal{C}_{\mathrm{edit}})$. Repeating the text enforces content preservation, while the shared speaker condition encourages identity consistency. During refinement, the model is invoked in synthesis mode for initial generation or severe failures, and in edit mode for correctable expressive mismatches.

\subsubsection{Training Data and Objective}

Because naturally paired speech samples that share identical content and speaker identity but differ in specific expressive attributes are scarce, we adopt a two-stage training procedure. In the first stage, we continue training Qwen3-TTS-12Hz-1.7B-CustomVoice to strengthen instruction-following synthesis for emotion, speaking rate, and loudness at different intensity levels. We observe that the original model exhibits nontrivial controllability for out-of-domain speaker. We therefore randomly sample speaker conditions and synthesize speech using diverse control instructions. Each generated sample is screened to verify that the requested control is reflected in the resulting speech, and only successful samples are retained. This procedure yields approximately 2,000 hours of instruction-following synthesis data for the first-stage control model.

In the second stage, we construct paired examples for attribute editing. For each text and speaker condition, we sample different control instructions and intensity levels to produce distinct expressive realizations. Two samples that share the same text and speaker identity but differ in their target attributes are organized into a source--target pair. The source speech and its original instruction form the first-turn context, whereas the target instruction and corresponding target speech provide supervision for the second turn. We construct approximately 1,400 hours of such edit-training data and supplement it with approximately 300 hours of recorded speech to improve acoustic diversity.

To prevent the model from losing its original synthesis capability while learning to edit, the second-stage training corpus also includes synthesis examples randomly sampled from the first-stage data. Synthesis examples retain the original autoregressive audio-token prediction objective, whereas edit examples are supervised only on the target audio tokens in the second turn. Let $\mathcal{I}_{2}$ denote the target-token positions of an edit example. The edit objective is
\begin{equation}
\mathcal{L}_{\mathrm{edit}}
=
-\sum_{n \in \mathcal{I}_{2}}
\log p_{\theta}
\left(
y_n \mid y_{<n}, \mathcal{C}_{\mathrm{edit}}
\right).
\end{equation}
where $y_n$ is the target audio token at position $n$. The first turn serves exclusively as source context and is excluded from the prediction loss. By interleaving synthesis and edit examples during this stage, the model learns to generate the target realization from the source speech and edit instruction while retaining its instruction-following synthesis capability.

\subsection{Agentic Refinement Framework}

High-quality multi-speaker dialogue synthesis requires both utterance-level instruction following and dialogue-level consistency. An utterance may sound satisfactory in isolation but become inconsistent when placed in context. For example, two adjacent turns in an argument may both convey anger, yet differ sharply in intensity or speaking manner, resulting in an unnatural expressive transition. Such failures cannot be reliably addressed by independently synthesizing and concatenating individual utterances.

We therefore formulate dialogue synthesis as a closed-loop agentic refinement process consisting of structured planning, initial synthesis, hierarchical critique, and targeted revision. As illustrated in Figure~\ref{fig:method_overview}(a), the framework evaluates generated speech at both utterance and scene levels and translates critique feedback into executable refinement actions.

\subsubsection{Structured Planning and Initial Synthesis}

Given an input script, Gemini 2.5 Flash~\cite{gemini25} serves as the planning model and converts the unstructured content into a dialogue plan. For each turn, the plan specifies the utterance text, speaker identity, expressive instruction, and contextual information. The context summarizes character and scene states that may influence how an utterance should be realized. This structured representation makes implicit requirements in story-like inputs explicit and provides a shared conditioning interface for synthesis and critique.

The framework then processes the dialogue turn by turn. For each turn $i$, the text, speaker prompt, and control instruction are passed to the synthesis mode of ControlEdit-TTS to produce an initial speech segment $x_i$. The generated segment remains associated with its dialogue-plan metadata, allowing subsequent critique to evaluate it against both its local instruction and its role in the surrounding dialogue.

\subsubsection{Hierarchical Critique and Targeted Refinement}

The hierarchical critique mechanism operates first at the utterance level and then at the scene level. At the utterance level, Gemini 2.5 Pro~\cite{gemini25} evaluates instruction adherence, linguistic content and pronunciation, acoustic artifacts, and speaker identity. An utterance that satisfies these criteria is accepted. Otherwise, the critique module diagnoses the failure and determines an appropriate refinement action.

Failures involving editable expressive attributes, such as emotion, speaking rate, or loudness, are handled by invoking the edit mode of ControlEdit-TTS on the existing speech. Failures involving incorrect content, pronunciation errors, severe acoustic artifacts, or an incorrect speaker instead trigger resynthesis. The revised speech is returned to the utterance-level critique loop until it is accepted or the maximum number of refinement rounds is reached. Both actions are performed by the same ControlEdit-TTS model; the framework changes the task mode and input context rather than switching between separate generation and editing models.

After all turns pass the utterance-level critique, they are composed into a complete dialogue timeline. Gemini 3.1 Pro Preview then performs scene-level critique of speaker consistency across turns, transition smoothness, dialogue coherence, and contextual appropriateness. 
Scene-level critique instead evaluates the composed dialogue and identifies turns or boundaries responsible for cross-turn inconsistencies.

The resulting context-aware feedback is mapped to targeted operations. Temporal discontinuities are addressed by adjusting inter-utterance gaps, local expressive inconsistencies are corrected through editing, and severe or non-editable failures trigger selective resynthesis of the affected turns. The updated dialogue is recomposed and evaluated again, allowing the framework to refine only the problematic regions while preserving accepted speech elsewhere. To prevent cyclic correction, the refinement process is terminated after at most 10 rounds if unresolved issues remain.

Through this action-oriented critique policy, feedback from the speech understanding model is converted into concrete synthesis operations rather than serving only as an acceptance score. The resulting planning--synthesis--critique--refinement loop augments the underlying speech generator with dialogue-aware quality control, improving instruction adherence and cross-turn consistency in long-form multi-speaker dialogue synthesis.

\section{Experiments}
\begin{table*}[t]
\centering
\small
\setlength{\tabcolsep}{5pt}
\begin{tabular}{lcccccc}
\toprule
\multirow{2}{*}{Model} & \multicolumn{2}{c}{Speaker similarity$\uparrow$} & \multicolumn{4}{c}{Instruction following} \\
\cmidrule(lr){2-3} \cmidrule(lr){4-7}
& ZH & EN & Yes\%$\uparrow$ & Partial\% & No\%$\downarrow$ & Avg. score$\uparrow$ \\
\midrule
\multicolumn{7}{l}{\textit{Proposed system and ablations}} \\
ControlEdit-TTS (full) & 0.776 & 0.791 & \textbf{95.5} & 4.2 & 0.3 & \textbf{4.819} \\
ControlEdit-TTS (edit disabled) & 0.777 & 0.791 & 87.0 & 12.3 & 0.7 & 4.639 \\
ControlEdit-TTS (one-shot) & 0.752 & 0.776 & 85.2 & 13.8 & 1.0 & 4.592 \\
\midrule
\multicolumn{7}{l}{\textit{Agentic baselines with our framework}} \\
Higgs-Audio v3 & 0.709 & 0.790 & 88.8 & 10.5 & 0.7 & 4.689 \\
VoxCPM2  & 0.696 & 0.761 & 83.3 & 15.3 & 1.4 & 4.544 \\
\midrule
\multicolumn{7}{l}{\textit{Direct dialogue models}} \\
SoulX-Podcast & \textbf{0.827} & \textbf{0.861} & 75.0 & 22.6 & 2.5 & 4.321 \\
Fish Audio S2 & 0.736 & 0.780 & 78.2 & 19.7 & 2.1 & 4.404 \\
VibeVoice-7B & 0.763 & 0.790 & 81.1 & 17.2 & 1.8 & 4.484 \\
VibeVoice-1.5B & 0.748 & 0.782 & 78.5 & 19.5 & 2.0 & 4.408 \\
Any2Speech & 0.620 & 0.695 & 75.7 & 22.0 & 2.3 & 4.345 \\
\bottomrule
\end{tabular}
\caption{Utterance-level instruction following and speaker preservation. The automatic judge assigns Yes, Partial, or No and a score from 1 to 5. One-shot denotes the shared initial synthesis before critique-guided refinement.}
\label{tab:utterance_level_results}
\end{table*}

\subsection{Experimental Setup}
\paragraph{Evaluation benchmark.}

We construct a bilingual benchmark for controllable multi-speaker dialogue synthesis with 58 Chinese and 60 English stories. It covers short stories, scene scripts, and dialogue excerpts with diversity in genre, length, speaker count, and speaker age. Each story is converted into a structured sequence of speaker identity, expressive instruction, and utterance text. For each character, we create a global description and use Qwen3-TTS-12Hz-1.7B-VoiceDesign~\cite{Qwen3-TTS} to generate the reference voice.

\paragraph{Compared systems.}

We compare with five direct multi-speaker dialogue systems: SoulX-Podcast~\cite{soulxpodcast}, Fish Audio S2~\cite{fishs2}, VibeVoice-1.5B, VibeVoice-7B~\cite{vibevoice}, and Any2Speech~\cite{any2speech}. SoulX-Podcast and VibeVoice directly synthesize multi-speaker dialogue but do not support natural-language utterance-level style control. Fish Audio S2 supports inline tag-based control, and Any2Speech accepts structured dialogue scripts with speaker references through a commercial API. We use official checkpoints or APIs with
default inference settings.
We further build two agentic baselines by replacing our backbone with Higgs-Audio v3~\cite{higgs_audio_tts_v3} and VoxCPM2~\cite{voxcpm2}. Since these models do not support attribute editing, all editable failures are handled by resynthesis.

We evaluate four variants of ControlEdit-TTS: \textbf{Full}, the complete framework; \textbf{Edit disabled}, which replaces editing with resynthesis; \textbf{One-shot}, the shared initial synthesis before refinement; and \textbf{Utterance-level only}, which omits scene-level critique and refinement.

For direct dialogue baselines, we evaluate whether the generated utterance matches the target expressive requirement implied by the utterance semantics and dialogue context. For systems without explicit utterance-level controls, this requirement is assessed only from the generated output. For Fish Audio S2, which supports inline tag-based control, we express the target requirement using its inline control tags.

\paragraph{Implementation details.}
ControlEdit-TTS is initialized from Qwen3-TTS-12Hz-1.7B-CustomVoice and trained in two stages. In stage 1, we continue training on the 2,000-hour instruction-following corpus to strengthen controllable synthesis of emotion, speaking rate, and loudness. In stage 2, we further train the same model on a mixture of stage-1 synthesis data, the 1,400-hour synthetic edit corpus, and 300 hours of recorded speech to preserve synthesis capability while learning attribute editing. We use the AdamW optimizer with a peak learning rate of $1\times10^{-5}$ on 32 A100 GPUs, and train each stage until convergence.

At inference time, the planning module uses Gemini 2.5 Flash, while utterance-level and scene-level critique are performed by Gemini 2.5 Pro and Gemini 3.1 Pro Preview, respectively. All prompts are fixed across experiments. The refinement loop is capped at 10 rounds. Utterance-level refinement is performed turn by turn, and accepted utterances are then composed for scene-level critique. Expressive mismatches in emotion, speaking rate, or loudness are routed to edit mode, while content errors, pronunciation failures, severe artifacts, or speaker mismatches trigger resynthesis. Timing adjustment modifies only inter-utterance silence duration.

\paragraph{Evaluation protocol.}

We evaluate systems at both utterance and dialogue levels. At the utterance level, Gemini 2.5 Pro judges whether each utterance matches its target expressive requirement and returns a \textit{yes}, \textit{partial}, or \textit{no} label together with a 1--5 score. For controllable systems, this requirement is given by the utterance-level instruction; for direct dialogue systems, it is treated as the desired expressive style implied by the utterance and its context. Following recent work~\cite{Instructttseval,mintbench}, we use Gemini-based judgments as an automatic proxy for instruction adherence. For direct dialogue baselines that synthesize a whole dialogue in one pass, we first align the generated waveform to the target script with the open-source Qwen3-ForceAligner~\cite{qwen3asr,LLM-ForcedAligner} and extract utterance-level clips from the aligned boundaries. Speaker preservation is then measured as cosine similarity between Res2Net speaker embeddings~\cite{res2net} extracted from each utterance clip and the corresponding character reference voice.

At the dialogue level, Gemini 3.1 Pro Preview and human raters conduct pairwise preference evaluation on continuous multi-turn clips, considering role continuity, cross-turn expressive coherence, and overall listening quality. We also analyze residual voice and style issues across refinement rounds; agreement between Gemini-based and human dialogue judgments is reported in the appendix.

\subsection{Utterance-level Control and Speaker Preservation}
We first evaluate whether each synthesis configuration follows utterance-level expressive instructions while preserving the target speaker identity. Gemini 2.5 Pro evaluates every utterance in the bilingual test set using the utterance text and target instruction. Speaker similarity is computed against the reference voice assigned to the corresponding role. The one-shot output serves as the shared initial state, while the two refined configurations differ only in whether editable failures are corrected through attribute editing or resynthesis.

As shown in Table~\ref{tab:utterance_level_results}, one-shot synthesis achieves an 85.2\% Yes rate and a 4.592 average score. Critique-guided refinement with resynthesis alone improves these values modestly to 87.0\% and 4.639, while the full system further reaches 95.5\% and 4.819. These comparisons separate the contributions of iterative refinement and edit-based correction: the one-shot baseline shows the quality of initial generation, the edit-disabled variant shows the effect of critique-guided resynthesis, and the full system shows the additional benefit of targeted editing.

Relative to one-shot synthesis, the full framework improves the Yes rate by 10.3 percentage points and slightly increases speaker similarity in both languages. This SIM improvement is consistent with the critique loop, which can also detect speaker mismatches and trigger corrective resynthesis. In contrast, the full and edit-disabled variants exhibit nearly identical speaker similarity but differ by 8.5 percentage points in Yes rate, indicating that targeted editing is the main source of additional instruction-following gain beyond initial synthesis and resynthesis alone. Among all compared systems, ControlEdit-TTS (full) achieves the highest Yes rate and average score. Although SoulX-Podcast attains the highest reference-voice similarity, its much lower instruction-following performance suggests that preserving voice identity alone is insufficient for controllable dialogue generation. Fish Audio S2 improves over other direct dialogue baselines in expressive control, likely due to its inline control interface, but remains clearly below the proposed system.

\vspace{-5pt}
\subsection{Dialogue-Level Preference}

\begin{table}[t]
\centering
\small
\setlength{\tabcolsep}{1.7pt}
\begin{tabular}{lcc@{\hspace{4pt}}cc}
\toprule
\multicolumn{1}{c}{\multirow{2}{*}{Opponent}}
& \multicolumn{2}{c}{Chinese ($N=93$)}
& \multicolumn{2}{c}{English ($N=94$)} \\
\cmidrule(lr){2-3}\cmidrule(lr){4-5}
& W/L/T & WR 
& W/L/T & WR  \\
\midrule
\multicolumn{5}{l}{\textit{Direct dialogue systems}} \\
SoulX-Podcast
    & 81/12/0 & 0.871
    & 71/19/4 & 0.755 \\
Fish Audio S2
    & 73/19/1 & 0.785
    & 67/18/9 & 0.713 \\
VibeVoice-7B
    & 56/36/1 & 0.602
    & 59/26/9 & 0.628 \\
VibeVoice-1.5B
    & 76/17/0 & 0.817
    & 63/19/12 & 0.670 \\
Any2Speech
    & 68/25/0 & 0.731
    & 66/26/2 & 0.702 \\
\midrule
\multicolumn{5}{l}{\textit{Agentic baselines with our framework}} \\
Higgs-Audio v3
    & 67/25/1 & 0.720
    & 64/30/0 & 0.681 \\
VoxCPM2
    & 52/41/0 & 0.559
    & 70/23/1 & 0.745 \\
\midrule
\multicolumn{5}{l}{\textit{Ablation}} \\
Ours (utterance only)
    & 79/3/11 & 0.849
    & 82/4/8 & 0.872 \\
\bottomrule
\end{tabular}
\caption{Gemini pairwise preference for ControlEdit-TTS (full). W/L/T denote our wins/losses/ties, and $\mathrm{WR}=\mathrm{W}/(\mathrm{W}+\mathrm{L}+\mathrm{T})$. The utterance-only variant omits scene-level critique and refinement.}
\label{tab:pairwise_results}
\vspace{-15pt}
\end{table}
Utterance-level scores do not fully capture role continuity, expressive progression, or transition quality across turns. We therefore conduct pairwise evaluation on continuous clips sampled from the generated dialogues. Each clip contains at least six turns and three distinct speakers, with a total duration below one minute. The same sampling protocol is used for the Gemini-based and human evaluations.
\paragraph{Gemini-based pairwise evaluation.}
Table~\ref{tab:pairwise_results} reports the pairwise preference of ControlEdit-TTS (full) against every baseline on the Chinese and English test sets.
Table~\ref{tab:pairwise_results} shows that ControlEdit-TTS (full) is preferred over every external baseline in both languages, with win proportions ranging from 0.559 to 0.871 in Chinese and from 0.628 to 0.755 in English. The consistently positive margins indicate that the proposed framework improves overall dialogue quality not only against direct multi-speaker systems, but also against strong controllable TTS backbones embedded in the same refinement framework. Among direct dialogue baselines, Fish Audio S2 is more competitive than SoulX-Podcast and Any2Speech in controllability-oriented comparisons, but remains consistently behind the proposed system.

The scene-level ablation further reveals a substantial contribution from contextual refinement. Compared with the utterance-level-only variant, the full system achieves win proportions of 0.849 in Chinese and 0.872 in English. Since the two configurations are identical through planning, initial synthesis, and utterance-level refinement, this comparison isolates the value of scene-level critique and its subsequent context-aware editing, resynthesis, and timing adjustment. The result shows that dialogue-level consistency cannot be recovered reliably from utterance-level correction alone.

\paragraph{Human subjective evaluation.}

To complement the automatic evaluation, we conduct a human pairwise preference test with five raters. For each opponent, 30 dialogue pairs are evaluated, and each pair is independently rated by three raters, yielding 90 individual judgments per comparison. The presentation order of the two systems is randomized for every trial. Raters compare the complete multi-turn clips in terms of role continuity, cross-turn expressive coherence, transition naturalness, and overall listening quality. A tie is allowed when neither system is clearly preferred.

\begin{figure}[t]
    \centering
    \includegraphics[width=\columnwidth]{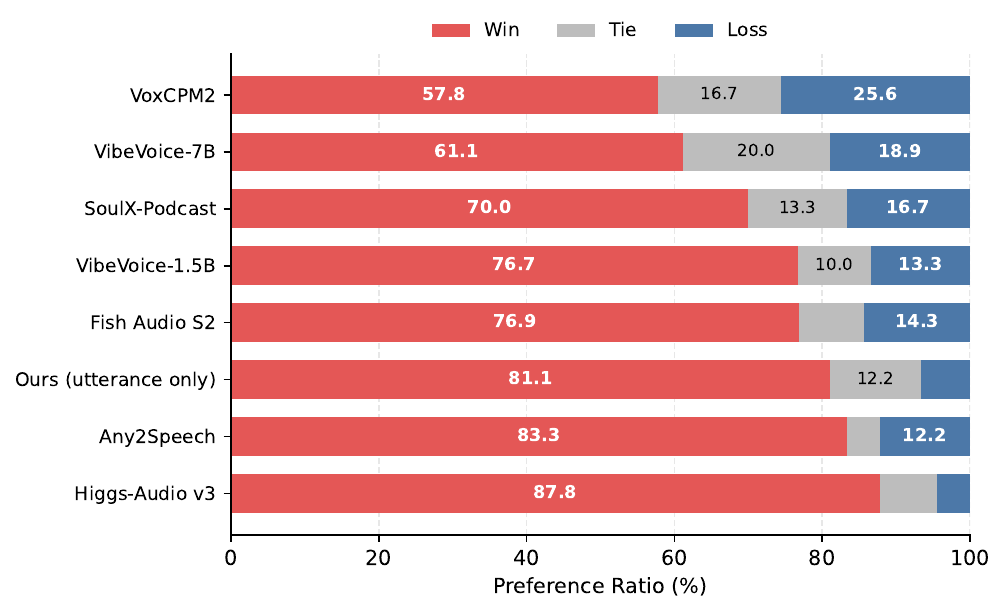}
    \caption{Human pairwise preference results for ControlEdit-TTS (full). Each horizontal bar shows the proportions of wins, ties, and losses against one opponent, sorted by win ratio. ``Ours (utterance only)'' denotes the ablation without scene-level critique and refinement.}
    \label{fig:human_ab}
    \vspace{-10pt}
\end{figure}

As shown in Figure~\ref{fig:human_ab}, ControlEdit-TTS (full) is preferred over every external baseline, with win proportions ranging from 57.8\% against VoxCPM2 to 87.8\% against Higgs-Audio v3. The largest human preference margins are observed against Higgs-Audio v3 and Any2Speech, while VoxCPM2 remains the closest external competitor. Fish Audio S2 is more competitive than most direct dialogue baselines, likely due to its inline tag-based control, but the proposed system is still preferred in 76.9\% of judgments. Importantly, the full system is also preferred over the utterance-level-only ablation, receiving 81.1\% wins against only 6.7\% losses. Because the two variants differ only in the presence of scene-level critique and refinement, this result provides direct human evidence that dialogue-level contextual correction improves multi-turn listening quality beyond utterance-level control alone. The overall preference trend is consistent with the Gemini-based evaluation.

\subsection{Effectiveness of Edit-based Refinement}

We isolate the effect of editing within the refinement process. ControlEdit-TTS (full) resolves editable attribute mismatches through editing, whereas ControlEdit-TTS (edit disabled) handles the same cases with resynthesis; severe failures trigger resynthesis in both settings. We select 17 challenging benchmark cases that exhibit frequent refinement problems and are therefore harder than average to synthesize reliably, using them to stress-test correction dynamics.
For each case, we record the shared initial state and then perform five refinement rounds. After the initial state and each round, the critique module counts remaining \textit{voice issues}, defined as within-role speaker inconsistencies, and \textit{style issues}, defined as mismatches with the target expressive instruction. Results are averaged per dialogue, and the total count is their sum.

\begin{figure}[t]
    \centering
    \includegraphics[width=0.75\columnwidth]{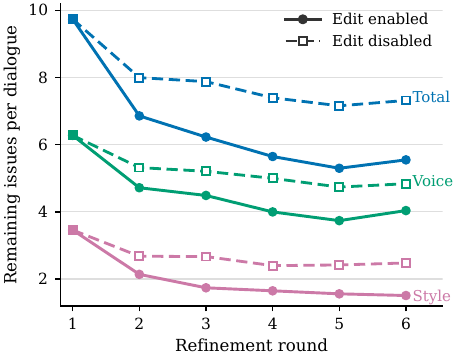}
    \caption{Average number of remaining issues per dialogue. The first point is the shared initial state before
    correction, followed by five refinement rounds. Solid lines with filled circles
    denote ControlEdit-TTS (full), whereas dashed lines with open
    squares denote ControlEdit-TTS (edit disabled). Lower is better.}
    \label{fig:refinement_efficiency}
    \vspace{-10pt}
    
\end{figure}

Figure~\ref{fig:refinement_efficiency} shows that edit-based refinement reduces residual issues more rapidly than regeneration alone. By round 4, ControlEdit-TTS (full) reduces the average total issue count from 9.74 to 5.30, corresponding to a 45.6\% reduction, whereas the edit-disabled configuration reaches 7.16 issues, corresponding to a 26.5\% reduction. At the final round, the full system also retains fewer style and voice issues. The larger improvement in style correction supports the use of localized editing for resolving expressive mismatches without regenerating the complete utterance.

Both configurations exhibit a small rebound in total and voice issue counts from round 4 to round 5. For the full system, the total count increases from 5.30 to 5.55 as the voice issue count rises from 3.74 to 4.04. Case-level analysis suggests that this rebound often reflects a trade-off between style correction and voice consistency: improving style alignment may introduce speaker inconsistency, whereas correcting speaker-related issues may weaken style alignment. This seesaw behavior indicates that the current synthesis/editing model still has limited capacity to jointly satisfy these constraints under repeated correction. It also suggests diminishing returns from prolonged refinement and motivates stronger backbone modeling and adaptive stopping strategies. Overall, edit-based correction achieves faster convergence and consistently fewer residual issues than regeneration alone.

\section{Conclusion}
We presented a framework for controllable multi-speaker dialogue TTS that formulates synthesis as a critique-driven iterative refinement process rather than one-shot generation. The system combines structured planning, turn-level synthesis, hierarchical utterance- and scene-level critique, and targeted revision. Its speech backbone, ControlEdit-TTS, unifies instruction-following synthesis and attribute-level editing of emotion, speaking rate, and loudness within a single autoregressive model. Experiments on a bilingual Chinese--English benchmark show improved utterance-level instruction following, stronger dialogue-level preference, and more effective refinement than regeneration-only alternatives, while preserving speaker identity.
Our results suggest that controllable dialogue speech generation benefits from tightly integrating speech generation and speech understanding in a closed-loop process. Future work will extend the editable attribute space, improve robustness in longer and more complex scenes, and develop stronger critique and stopping strategies for more efficient refinement.

\bigskip

\bibliography{aaai2027}

\clearpage
\onecolumn
\raggedbottom

\setcounter{section}{0}
\setcounter{subsection}{0}
\setcounter{subsubsection}{0}
\setcounter{table}{0}
\setcounter{figure}{0}
\setcounter{equation}{0}
\renewcommand{\thesection}{S\arabic{section}}
\renewcommand{\thesubsection}{S\arabic{section}.\arabic{subsection}}
\renewcommand{\thesubsubsection}{S\arabic{section}.\arabic{subsection}.\arabic{subsubsection}}
\renewcommand{\thetable}{S\arabic{table}}
\renewcommand{\thefigure}{S\arabic{figure}}
\renewcommand{\theequation}{S\arabic{equation}}
\setcounter{secnumdepth}{3}

\begin{center}
{\Large\bfseries Supplementary Material for\par}
\vspace{0.5em}
{\large\bfseries From Script to Drama: An Agentic Framework for Controllable Multi-Speaker Dialogue TTS\par}
\vspace{0.8em}
{\normalsize Kangxiang Xia et al.\par}
\end{center}

\vspace{1em}

\noindent
This supplementary material provides additional dataset, pipeline,
evaluation, and baseline details omitted from the main paper because of
space constraints. We first describe the construction of the bilingual
controllable dialogue TTS benchmark, including its sampling procedure,
structured annotations, and distributional statistics. We then summarize the
multi-stage synthesis pipeline and the prompts used in script parsing,
single-utterance quality control, scene-level critique, and instruction
mapping. Finally, we provide additional human-evaluation details and raw
counts used in the main paper.

\section{Benchmark Construction Details}
\label{sec:supp_benchmark}

We construct a bilingual benchmark for controllable multi-speaker dialogue TTS,
consisting of 57 Chinese stories and 60 English stories. The benchmark is
designed for multi-character long-form dialogue generation with voice cloning
and utterance-level expressive control. It contains short stories, scene
scripts, and dialogue-centered narrative excerpts, and explicitly evaluates
both local utterance realization and dialogue-level coherence.

Unless otherwise stated, all statistics and prompt descriptions in this
appendix follow the \textbf{speaker-only dialogue-line} protocol: only spoken
dialogue lines and their associated speaking characters are considered.

\subsection{Construction Pipeline}
\label{sec:supp_pipeline}

The Chinese and English test sets are built using a mirrored construction
pipeline, differing only in language and localized genre definitions. The
overall procedure consists of four stages: (1) sampling narrative seeds under
controlled distributions of genre, speaker count, target length, emotional
intensity, and character demographics; (2) generating candidate scripts from
multiple large language models and selecting one candidate per seed; (3)
parsing the selected story into a structured dialogue script with line-level
expressive annotations; and (4) assigning reference voices to all speaking
characters.

For both languages, we define 15 genre categories and impose balancing
constraints so that the benchmark covers a diverse range of narrative
situations while preventing a small number of difficult genres from dominating
the test set.

\subsection{Seed Sampling}
\label{sec:supp_seed_sampling}

We first sample 60 narrative seeds for each language. Each seed
specifies a genre, target speaker count, target story length, emotional
intensity, and character demographic configuration.

The sampling rules are as follows. Genres are drawn from 15 predefined
categories, with upper bounds imposed on several difficult genres so that they
do not dominate the benchmark. In Chinese, these capped categories are
historical-period, wuxia/xianxia, sci-fi/future, and supernatural/ghost-story
scenarios; in English, they are Historical, High Fantasy, Sci-Fi, and Horror.
Target speaker count follows a quota distribution of 30\% two-speaker cases,
30\% three-speaker cases, 20\% four-speaker cases, and 20\% five-speaker
cases. Target length follows a quota distribution of 40\% short, 40\% medium,
and 20\% long cases. Emotional intensity is sampled between relatively flat and
strongly varying scenes at approximately equal proportions. Character gender
and age are diversified through a biasing scheme that first covers all six
gender--age groups before reusing them.

\subsection{Story Generation and Selection}
\label{sec:supp_story_generation}

For each sampled seed, two large language models independently generate a
candidate story script: Gemini 2.5 Pro and GPT-5.4. We then select the version
with fewer dialogue turns; ties are further broken by the total number of
dialogue lines, and remaining ties are resolved in favor of Gemini. This
procedure controls test-set length while preserving diverse narrative
structure. Because GPT-5.4 tends to produce longer scripts on average, the
final selected pool is Gemini-dominant in both languages.

\subsection{Structured Dialogue Scripts}
\label{sec:supp_dialogue_plan}

Each selected story is converted into a structured dialogue script stored as
\texttt{script.json}. The script contains two components: a set of character
entries and a sequence of dialogue lines. Each character entry includes a
character identifier, a name, a voice description, a reference waveform, and
the corresponding reference transcript. Each dialogue line includes its speaker
identity, utterance text, and a set of expressive-control fields, including
emotion, speaking rate, loudness, post-utterance pause, instruction text, and
scene description.

More concretely, the parser produces:
\begin{itemize}
    \item \texttt{characters[]}: one entry per speaking character, including
    \texttt{id}, \texttt{name}, \texttt{voice\_description},
    \texttt{reference\_wav}, \texttt{reference\_text}, and
    \texttt{voice\_mode=clone};
    \item \texttt{lines[]}: one entry per dialogue line, including
    \texttt{index}, \texttt{character\_id}, \texttt{text}, and structured
    control labels such as \texttt{emotion}, \texttt{pace},
    \texttt{volume}, \texttt{instruct}, \texttt{pause\_after}, and
    \texttt{scene\_desc}.
\end{itemize}

Each speaking character is paired with one reference recording of
approximately 8 seconds at 24~kHz, together with its transcript, for zero-shot
voice cloning.

\subsection{Directory Structure}
\label{sec:supp_directory}

The benchmark follows the directory organization below:
\begin{verbatim}
testset/
|-- manifest.jsonl / manifest.md
|-- runs/
|   `-- seed_XXX/
|       |-- script.json
|       |-- voices/<id>_ref.wav
|       |-- stitch_manifest.json
|       `-- ...
`-- runs_<engine>/
\end{verbatim}

Here, \texttt{manifest.jsonl} and \texttt{manifest.md} store case-level
metadata such as dialogue-turn counts, speaker counts, genre, target length,
and source model. Each case directory contains the parsed script, one
reference waveform per speaker, and synthesis artifacts such as
\texttt{stitch\_manifest.json} and the final rendered waveform. The
\texttt{runs/} directory corresponds to the main synthesis pipeline, whereas
\texttt{runs\_<engine>} stores outputs of other TTS engines on the same
scripts and reference voices for horizontal comparison.

\subsection{Benchmark Statistics}
\label{sec:supp_benchmark_stats}

Table~\ref{tab:supp_benchmark_overview} summarizes the benchmark at the case
level. The English set contains more dialogue lines overall and denser
dialogue per case, whereas the Chinese set contains slightly more speakers per
case and longer target story length in terms of raw text.

\begin{center}
\small
\captionof{table}{Overall benchmark statistics under the speaker-only dialogue-line protocol.}
\label{tab:supp_benchmark_overview}
\begin{tabular}{lcc}
\toprule
Statistic & Chinese & English \\
\midrule
\# cases & 57 & 60 \\
Dialogue lines & 2,309 & 3,196 \\
Dialogue lines / case & 40.5 (18--87) & 53.3 (10--152) \\
Speakers / case & 3.6 (2--7) & 3.4 (2--5) \\
Reference voice recordings & 206 & 201 \\
Mean reference duration & 8.0~s & 7.9~s \\
Reference duration range & 3.7--14.7~s & 3.6--13.3~s \\
Sampling rate & 24~kHz & 24~kHz \\
Target length & 1500--6000 chars & 600--2500 words \\
Source model (Gemini / GPT-5.4) & 51 / 6 & 57 / 3 \\
\bottomrule
\end{tabular}
\end{center}

\vspace{8pt}

Table~\ref{tab:supp_case_scale} reports per-case scale statistics. Chinese
cases contain fewer dialogue lines on average than English cases, but slightly
more speakers.

\begin{center}
\small
\captionof{table}{Per-case scale statistics.}
\label{tab:supp_case_scale}
\begin{tabular}{lcc}
\toprule
Statistic & Chinese & English \\
\midrule
Dialogue lines / case & mean 40.5, median 39, range 18--87 &
mean 53.3, median 45.5, range 10--152 \\
Speakers / case & mean 3.6, median 4, range 2--7 &
mean 3.4, median 3, range 2--5 \\
\bottomrule
\end{tabular}
\end{center}

\vspace{8pt}

Table~\ref{tab:supp_speaker_distribution} reports the distribution of
\emph{actual} speaker counts observed in the generated scripts. The
\texttt{num\_speakers} field in \texttt{manifest.jsonl} records the sampling
target (2--5 speakers), whereas the statistics below are computed from the
actual parsed script and may be larger when additional side characters are
introduced by the generation model.

\begin{center}
\small
\captionof{table}{Distribution of actual speaker counts in the structured scripts.}
\label{tab:supp_speaker_distribution}
\begin{tabular}{lcc}
\toprule
Speaker count & Chinese & English \\
\midrule
2 & 10 & 16 \\
3 & 17 & 19 \\
4 & 18 & 13 \\
5 & 10 & 12 \\
6 & 1 & 0 \\
7 & 1 & 0 \\
\bottomrule
\end{tabular}
\end{center}

\vspace{8pt}

Table~\ref{tab:supp_length_distribution} reports the target length distribution.

\begin{center}
\small
\captionof{table}{Target length distribution.}
\label{tab:supp_length_distribution}
\begin{tabular}{lcc}
\toprule
Length & Chinese & English \\
\midrule
Short & 24 & 24 \\
Medium & 24 & 24 \\
Long & 9 & 12 \\
\bottomrule
\end{tabular}
\end{center}

\vspace{8pt}

Table~\ref{tab:supp_genre_distribution} shows the genre distribution. The two
sets use mirrored genre definitions with localized naming, and both maintain a
broad spread over everyday, dramatic, speculative, and high-difficulty
narrative categories.

\begin{center}
\small
\captionof{table}{Genre distributions for the Chinese and English test sets.}
\label{tab:supp_genre_distribution}
\begin{tabular}{lclc}
\toprule
Chinese genre & Count & English genre & Count \\
\midrule
Urban workplace & 6 & Workplace / Corporate & 7 \\
Mystery / Detective & 6 & Coming-of-Age / Campus & 7 \\
School / Youth & 6 & Mystery / Detective & 6 \\
Business negotiation & 4 & Medical Drama & 5 \\
Medical emergency & 4 & Business / Negotiation & 4 \\
Romance & 4 & Romance & 4 \\
Supernatural / Ghost story & 4 & Horror / Supernatural & 4 \\
Historical period & 4 & Historical Period Drama & 4 \\
Travel encounter & 4 & Travel / Strangers & 4 \\
Sci-fi / Future & 4 & Sci-Fi / Future & 4 \\
Family drama & 3 & Family Drama & 3 \\
Food / Daily life & 3 & Food / Culinary & 3 \\
Street / Slice of life & 2 & Slice of Life & 2 \\
Legal / Courtroom & 2 & Legal / Courtroom & 2 \\
Wuxia / Xianxia & 1 & High Fantasy & 1 \\
\bottomrule
\end{tabular}
\end{center}

\subsection{Line-Level Expressive Annotations}
\label{sec:supp_annotations}

The benchmark includes dense line-level expressive annotations. Every dialogue
line is assigned an instruction field and a scene description, together with
structured labels for speaking rate, loudness, post-utterance pause, and
emotion. The Chinese and English sets both contain 11 emotion categories.

Table~\ref{tab:supp_control_distribution} summarizes the distributions of
speaking rate, loudness, and post-utterance pause over the dialogue lines
under the speaker-only dialogue-line protocol.

\begin{center}
\small
\captionof{table}{Distribution of structured line-level control annotations over dialogue lines.}
\label{tab:supp_control_distribution}
\begin{tabular}{llcc}
\toprule
Field & Label & Chinese & English \\
\midrule
\multirow{3}{*}{Speaking rate}
& medium & 1,166 & 1,794 \\
& slow & 842 & 977 \\
& fast & 301 & 425 \\
\midrule
\multirow{3}{*}{Loudness}
& normal & 1,622 & 2,074 \\
& quiet & 455 & 883 \\
& loud & 213 & 239 \\
\midrule
\multirow{4}{*}{Pause after}
& short & 968 & 2,080 \\
& medium & 909 & 409 \\
& long & 399 & 688 \\
& scene change & 33 & 19 \\
\bottomrule
\end{tabular}
\end{center}

Both languages also provide near-complete free-form instruction and scene
description annotations: \texttt{instruct} is present for 100\% of Chinese
dialogue lines and 99.97\% of English dialogue lines, while
\texttt{scene\_desc} is present for 99.96\% of Chinese dialogue lines and
100\% of English dialogue lines.

The emotion distributions are as follows. In Chinese, the 11 emotion
categories are distributed as: calm 556, gravity 315, tenderness 300,
tension 205, coldness 182, anger 177, sadness 175, ease 145, joy 123,
surprise 79, and fear 52. In English, the 11 categories are distributed as:
gravity 549, calm 547, coldness 423, anger 357, sadness 356, tenderness 278,
ease 198, tension 195, fear 129, joy 92, and surprise 72.

These distributions ensure that the benchmark evaluates not only multi-speaker
alternation but also fine-grained utterance-level expressive control under
diverse dialogue conditions.

\subsection{Cross-Lingual Comparison}
\label{sec:supp_comparison}

The Chinese and English subsets are structurally parallel: both follow the same
pipeline, the same speaker-count and length sampling rules, mirrored genre
taxonomies, and the same line-level control schema for zero-shot cloned
dialogue TTS. The main differences are that the English set contains more cases
and denser dialogue per case, while the Chinese set contains slightly more
speakers per case and higher target story length in raw text. English
\texttt{pause\_after} labels are dominated by short pauses, whereas the Chinese
set shows a more balanced distribution between short and medium pauses.

\section{Synthesis Pipeline and Prompt Templates}
\label{sec:supp_prompts}

This section summarizes the multi-stage synthesis pipeline and the prompts used
by the LLM-based components. The pipeline consists of five stages:
script parsing, utterance synthesis, single-utterance quality control, signal
processing for stitching, and scene-level quality control with targeted
revision. 

\subsection{Pipeline Overview}
\label{sec:supp_pipeline_overview}

Table~\ref{tab:supp_pipeline_overview} summarizes the stages and their
associated prompts.

\begin{center}
\small
\captionof{table}{Pipeline stages and associated LLM prompts.}
\label{tab:supp_pipeline_overview}
\begin{tabularx}{\textwidth}{l l X}
\toprule
Stage & Role & LLM prompt(s) \\
\midrule
Stage 1 & Story text $\rightarrow$ structured \texttt{script.json} & Character extraction; line parsing \\
Stage 2 & Utterance-level TTS synthesis & No direct LLM prompt; uses instruction mappers \\
Stage 3 & Single-utterance acoustic QC and candidate selection & Single-utterance QC prompt \\
Stage 4 & Audio stitching, pause insertion, normalization & No prompt (signal processing only) \\
Stage 5 & Scene-level QC and targeted re-synthesis & Scene-level QC prompt; suggestion-to-key classification prompt \\
\bottomrule
\end{tabularx}
\end{center}

\subsection{Stage 1: Character Extraction Prompt}
\label{sec:supp_prompt_characters}

The first Stage-1 prompt reads the source story and extracts the set of
characters, together with one voice description and one reference-text snippet
for each character. The English version of the prompt is reproduced below.

\begin{quote}\small
You are a professional audiobook producer. Please read the following
audiobook text and extract all characters that appear, including the narrator.

Text:

\{text[:6000]\}

Please output a JSON array. Each element should contain:

- id: a unique English identifier for the character. The narrator must use
``narrator''. Other characters should use pinyin or an English-style short
identifier (e.g., ``lin\_mo'', ``gu\_ning'')

- name: the character's name in Chinese; for the narrator, use ``Narrator''

- voice\_description: a voice description suitable for this character
(gender, perceived age, timbre characteristics), for use in TTS voice design
mode; within 20 words

- voice\_mode: always set to ``design''

- reference\_wav: always set to null

- reference\_text: write a sample text snippet for generating this character's
reference audio (1--2 sentences, 20--40 Chinese characters). The content
should fit the character's personality, sound natural, and clearly reveal the
intended vocal qualities. For the narrator, use descriptive narrative text.
For dialogue characters, use conversational wording that matches the
character's personality.

Return only the JSON array. Do not provide any additional explanation.
\end{quote}

This prompt is used only during benchmark/script construction and does not
directly participate in inference-time synthesis.

\subsection{Stage 1: Line Parsing Prompt}
\label{sec:supp_prompt_parse_lines}

The second Stage-1 prompt parses one story chunk into structured dialogue-line
records and assigns line-level control fields such as \texttt{emotion},
\texttt{pace}, \texttt{volume}, \texttt{instruct}, \texttt{pause\_after}, and
\texttt{scene\_desc}. The English version of the prompt is reproduced below.

\begin{quote}\small
You are a professional audiobook script editor. Please parse the following
audiobook text segment (chunk \{chunk\_no\}/\{total\_chunks\}) into a sequence
of lines.

Known character list:

\{char\_list\_str\}

Text segment:

\{chunk\_text\}

Please parse every line of dialogue or narration in the text into one record.
Rules:

1. Narrative text (non-direct speech) should be assigned to narrator.

2. Direct speech inside quotation marks should be assigned to the speaking
character.

3. Each record must contain the following fields:

- scene\_desc: a short description of the scene/context for this line
(optional, within 10 words); do not include speaking style

- emotion: choose exactly one from the following labels:
calm, joy, sadness, anger, fear, tension, tenderness, coldness,
surprise, gravity, ease

- pace: choose one of: slow, medium, fast

- volume: choose one of: quiet, normal, loud

- instruct: a TTS reading instruction (within 15 words) that only describes
how the line should be spoken, such as ``low and slow, gently narrative'' or
``angry challenge, firm articulation''. Do not include scene description.

4. Choose \texttt{pause\_after} according to context:
short (continuous lines in the same scene), medium (topic shift),
long (end of paragraph), scene\_change (scene or time transition)

5. The line index should start from \{start\_index\} and increase
incrementally.

6. If the text contains a speaker who has already appeared but is not in the
character list, assign a suitable id following the naming style of the
existing characters.

Return only a JSON array. Do not provide any additional explanation.
\end{quote}

\subsection{Stage 3: Single-Utterance QC Prompt}
\label{sec:supp_prompt_single_qc}

Stage 3 evaluates one synthesized utterance at a time and checks whether it
follows the expected speaking style, sounds natural, and has acceptable audio
quality. The English version of the prompt is reproduced below.

\begin{quote}\small
You are a professional speech synthesis quality reviewer. Please carefully
listen to this synthesized speech and evaluate the following three aspects
from an acoustic perspective.

Expected speaking style: ``\{instruct\}''

1. style\_followed: Does the speaking style of the audio (speed, emotion,
tone) follow the instruction above?

2. naturalness\_ok: Does the audio sound natural and fluent?
(No obvious robotic quality, stuttering, unnatural phrasing, incomplete
sentence, or abnormal truncation.)

3. audio\_quality\_ok: Is the audio quality acceptable?
(No obvious noise, popping, clipping, abnormal silence, or overly short
duration.)

Output JSON only. Do not provide any additional explanation:

\{"style\_followed": true, "naturalness\_ok": true, "audio\_quality\_ok": true,
"issues": ["briefly describe issues if any"], "suggestion": "if re-synthesis is needed, describe a revised speaking-style instruction"\}
\end{quote}

\subsection{Stage 5: Scene-Level QC Prompt}
\label{sec:supp_prompt_scene_qc}

Stage 5 evaluates stitched multi-speaker audio segments together with the
corresponding numbered dialogue lines. It identifies content, voice, style,
and pause issues, and localizes them to specific line indices. The English
version of the prompt is reproduced below.

\begin{quote}\small
You are a professional audiobook production director. Please listen to this
multi-speaker dialogue audio. The corresponding lines are:

\{lines\_text\}

The numbers in square brackets are the line indices. When reporting problems,
you must use these indices to specify which lines are problematic.

Please evaluate the following aspects and identify the problematic lines:

1. Content accuracy: missing words, incorrect words, repetition, or
mispronunciation

2. Whether speaker transitions are clear and natural

3. Whether the pause rhythm is appropriate (too short or too long)

4. Whether the same character's voice remains consistent across lines

5. Whether the emotional expression matches the line content

Important: only mark problems that clearly affect the listening experience.
Minor and natural variation in timbre or tone is part of normal performance.
Do not over-label---a high-quality audio segment may have no problems at all
(output an empty array).

For each detected issue, output one record. The \texttt{seq} field must use the
line index shown in square brackets above. If there are no problems, output an
empty array.

Issues are divided into four types:

- type ``re\_synthesize'' + issue\_category ``content'': content errors.
Leave \texttt{suggestion} empty.

- type ``re\_synthesize'' + issue\_category ``voice'': timbre / audio quality /
pitch problems. Leave \texttt{suggestion} empty.

- type ``re\_synthesize'' + issue\_category ``style'': emotion / speaking rate /
loudness / tone problems. Write \texttt{suggestion} as one progressive English
instruction.

- type ``adjust\_pause'': inter-line pause is too short or too long. Use
\texttt{new\_pause} to specify the revised pause.

Output JSON only. Do not provide any additional explanation.
\end{quote}

\subsection{Suggestion-to-Key Classification Prompt}
\label{sec:supp_prompt_key_classification}

For edit-based targeted refinement, Stage 5 maps free-form English style
suggestions into one closed-set control key. The English version of the prompt
is as follows:

\begin{quote}\small
Classify the following English ``speaking style suggestion'' into the single
best-matching control key. You must choose only from the provided list. If no
key matches well, output \texttt{neutral}.

Available keys: \{\_PROG\_KEYS\}, neutral

Output exactly one key and nothing else.

Suggestion: ``\{suggestion\}''
\end{quote}

The candidate key set is drawn from the closed set of progressive control keys,
including emotion, speaking-rate, and loudness adjustment keys, plus
\texttt{neutral}.

\subsection{Non-LLM Stages}
\label{sec:supp_non_llm}

Several pipeline stages do not involve direct LLM prompting. Stage 2 performs utterance-level synthesis using the selected TTS backend. VoxCPM2 directly consumes the Chinese \texttt{instruct} field, whereas other backends may use backend-specific instruction preprocessing. Stage 4 performs only signal
processing, including clip stitching, pause insertion based on
\texttt{pause\_after}, and peak normalization. Finally, the story-generation
prompts used to write the underlying novel scripts belong to benchmark
construction rather than synthesis-time inference, and are therefore not part
of the online synthesis pipeline summarized here.

\section{Additional Experimental Results}
\label{sec:supp_results}

\subsection{Human Preference Raw Counts}
\label{sec:supp_human_raw}

Table~\ref{tab:supp_human_raw} reports the raw human pairwise preference counts
corresponding to the human pairwise preference figure in the main paper. Each
comparison aggregates 90 individual judgments, obtained from 30 dialogue pairs
with 3 raters per pair.

\begin{center}
\small
\captionof{table}{Raw human pairwise preference counts for ControlEdit-TTS (full).}
\label{tab:supp_human_raw}
\begin{tabular}{lccc}
\toprule
Opponent & Win & Tie & Loss \\
\midrule
SoulX-Podcast & 63 & 12 & 15 \\
VibeVoice-7B & 55 & 18 & 17 \\
VibeVoice-1.5B & 69 & 9 & 12 \\
Any2Speech & 75 & 4 & 11 \\
Higgs-Audio v3 & 79 & 7 & 4 \\
VoxCPM2 & 52 & 15 & 23 \\
Ours (utterance only) & 73 & 11 & 6 \\
Fish Audio S2 & 70 & 8 & 13 \\
\bottomrule
\end{tabular}
\end{center}

\section{Human Evaluation Details and Agreement}
\label{sec:supp_agreement}

\subsection{Human Preference Evaluation Protocol}
\label{sec:supp_human_details}

Five raters participated in the human pairwise preference evaluation. For each
opponent, we sample 30 dialogue pairs, and each pair is independently rated by
3 raters, resulting in 90 individual judgments per comparison. The order of
the two systems is randomized in every trial. Raters compare complete
multi-turn clips and select the preferred system based on role continuity,
cross-turn expressive coherence, transition naturalness, and overall listening
quality. A tie is allowed when neither system is clearly preferred.

For reporting in the main paper, pairwise results are aggregated as win, tie,
and loss counts over all individual judgments. The main paper visualizes the
corresponding proportions, and Table~\ref{tab:supp_human_raw} reports the raw
counts.

\subsection{Agreement Between Automatic Issue Detection and Human Verification}
\label{sec:supp_issue_agreement}

To assess the reliability of the automatic critique used in refinement
analysis, we compare automatically detected issues with human verification on a
sampled set of cases. We separately evaluate three categories: voice issues,
style issues (emotion, speaking rate, and loudness), and pause issues. For
each category, we count automatically flagged instances that are confirmed by
human inspection, false positives, and the resulting agreement rate.

Table~\ref{tab:supp_issue_agreement} shows that agreement is high across all
categories, reaching 0.902 for voice issues, 0.919 for style issues, and 0.841
for pause issues, with an overall agreement rate of 0.895. These results
support the use of the automatic critique module for issue counting in the
refinement analysis reported in the main paper.

\begin{center}
\small
\captionof{table}{Agreement between automatic issue detection and human verification.}
\label{tab:supp_issue_agreement}
\begin{tabular}{lccc}
\toprule
Issue type & Confirmed issues & False positives & Agreement \\
\midrule
Voice & 222 & 24 & 0.902 \\
Style (emotion/rate/loudness) & 79 & 7 & 0.919 \\
Pause & 58 & 11 & 0.841 \\
\midrule
Overall & 359 & 42 & 0.895 \\
\bottomrule
\end{tabular}
\end{center}

\end{document}